\documentclass[twocolumn, superscriptaddress, amsmath,amssymb]{revtex4}

\usepackage{subfigure,dcolumn}
\usepackage[T2A,T1]{fontenc}
\usepackage[russian,english]{babel}
\usepackage{graphicx}

\usepackage{listings}
\begin{document}

\title{First evidence for chiral wobbling of triaxial nuclei}

\author{S.~Frauendorf}
\email{sfrauend@nd.edu}
\affiliation{Department of Physics and Astronomy, University of Notre Dame, Notre Dame, Indiana 46556,  USA}



\maketitle

\begin{figure}[!htb]
\includegraphics
  [width=\linewidth]
  {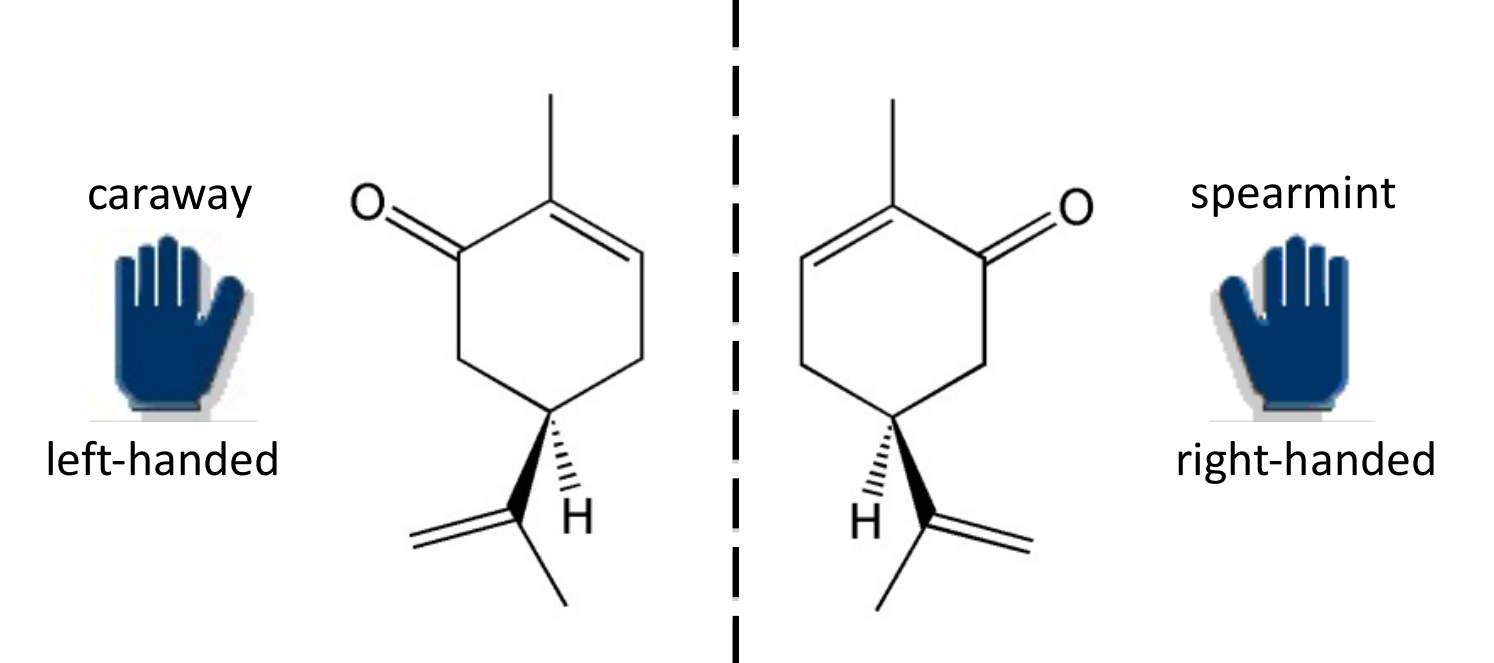}
\caption{\label{f:Carvon} The two chiral configurations of the carvon molecule. As usual in organic chemistry, the C atoms are not shown. They are located
at the bond ends.
 The left-handed molecule is transformed into the right-handed by a reflection through the mirror plane, which is   perpendicular to the figure plane
and goes through the broken line.
 }.
\end{figure}

Chirality (Greek "handiness") is a property of many complex molecules. Chiral molecules exist in two forms, one being the mirror image of the other. Like for our hands, it is impossible to make the images identical by a suitable rotation. The two forms are called left-handed and right-handed. They have the same binding energy, because the electromagnetic interaction, which holds the molecule together, does not change under a reflection. Other properties that are insensitive to the geometry are also the same. The different geometry is the reason why the left-handed form turns the polarization plane of transmitted light in one direction by some angle while of the right-handed form turns it in the opposite direction by the same angle. 
However, the geometrical differences between the two species may have other  consequences.  The two species of the carvon molecule shown in Figure \ref{f:Carvon} taste quite differently.

In contrast  to molecules, nuclei are very compact. They resemble deformed droplets of  liquid, with nearly constant interior density and a thin surface, which reflects the short range of the interaction that keeps the nucleons together.   The  shapes are simple: spherical, axial symmetric, mostly footbal$l$-like, sometimes pear-shaped. Some nuclei take the shape of a triaxial ellipsoid.
 Such simple shapes do not attain the property of chirality, which was considered as an alien concept by nuclear physicists.  Surprisingly, the studies of the rotational excitations of triaxial nuclei 
 have revealed that they may attain chirality \cite{FM97}.
 
\begin{figure}[!h]
\includegraphics
  [width=\linewidth]
  {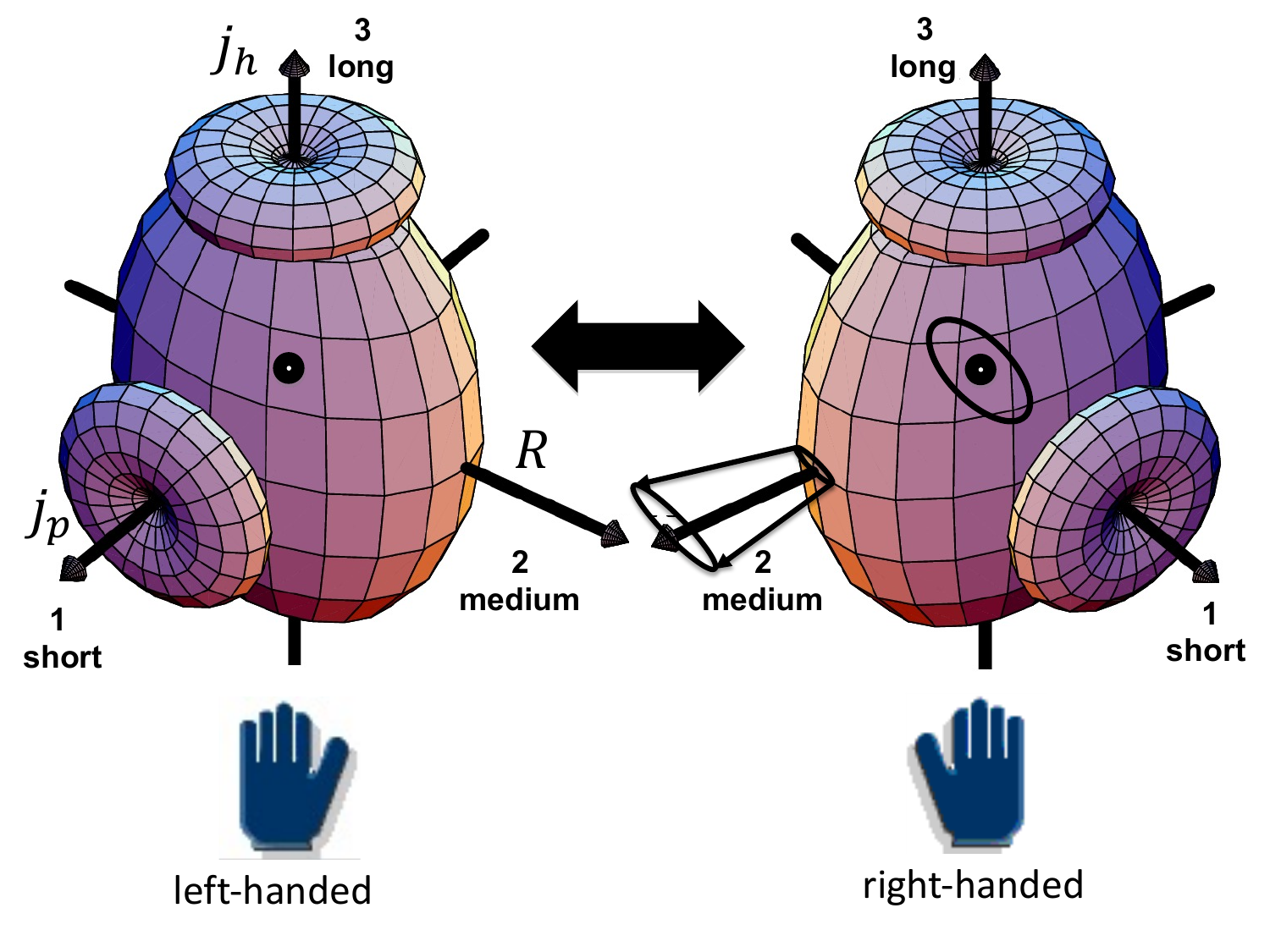}
\caption{A rotating triaxial nucleus with an unpaired particle carrying the angular momentum $\vec{j}_p$, an unpaired hole carrying $\vec{j}_h$ , and collective angular momentum $\vec{R}$ .  All angular momenta are displayed as arrows. The total angular momentum  $\vec{J}$, which is the
 axis of rotation,  points into the direction of sight and is depicted by the large dot symbolizing the arrow tip.
  For purpose of illustration the high-$j$ orbitals are drawn outside the nucleus, whereas they are located inside symmetric to the principle planes.
 The  cone of  $\vec{R}$ illustrates the excitation of the wobbling mode. 
 The ellipse around the dot  shows the precession of $\vec{J}$ for the chiral wobbling excitation.}
\label{f:shape}
\end{figure}

Deformed nuclei exhibit rotational spectra similar to molecules. These rotational bands are regular sequences of states with  quantized energy and angular 
momentum differing solely in their rotational velocity. Concerning the dynamics of rotation, molecules behave like classical bodies while nuclei behave like a quantum liquid.
In even-even nuclei, the rotational dynamics necessitates that the triaxial shape deviates from rotational symmetry with respect to a principal axis in order to facilitate rotation around  it. 
The corresponding moments of inertia increase with the asymmetry of the shape with respect to the axis. As illustrated by the triaxial shape in Fig. \ref{f:shape},
the asymmetry is maximal for the medium ($m$-) axis and so the moment of inertia. Therefore rotation about the $m$-axis generates the lowest (yrast) rotational band. Only even values of
the angular momentum $R$ appear because for a quantum liquid of indistinguishable constituents a rotation by the angle of $\pi$ leads to an identical state, which  restricts the
possible values of $R$ to even. 

In triaxial nuclei,  excited rotational bands appear when the angular momentum vector $\vec{R}$ precesses around the $m$-axis.
The energy of these "wobbling" excitations increases with the quantized opening angles of the precession cones, and the symmetry requires that $R$ is odd for the first wobbling
excitation, even for the second, etc.. Figure \ref{f:shape} shows the precession of  $\vec{R}$ with respect to the principal axes of triaxial shape. In the laboratory frame, $\vec{R}$ stands
still and the body executes a wobbling motion which counteracts the precession in the body-fixed frame. The accompanying motion of the charge quadrupole generates collectively
enhanced $E2$ transitions between the adjacent wobbling bands, which are the major evidence for  their wobbling nature. So far, the evidence for wobbling bands built on the yrast
band is restricted to a few even-even nuclei (see review \cite{Frau24}).

A new pattern evolves when  not all the angular momenta  $\vec{j}$ of the nucleonic orbitals are paired off, which appears in odd-$A$, odd-odd nuclei  and in bands built 
on quasiparticle  excitations in even-even nuclei.  Figure \ref{f:shape} illustrates  the shape of the high-$j$  orbitals, which, due to the spin-orbit splitting,
are $d_{5/2},~f_{7/2},~g_{9/2},~h_{11/2},~i_{13/2},~j_{15/2}$  to good approximation.  When an odd proton on  the doughnut-like orbital  is added to the even-even  
 "core" of the nucleus, the angular momentum of the orbital $\vec{j}_p$  points along the $s$-axis. 
 This orientation corresponds to the energy minimum of the short-range attractive interaction between the orbital and the core. 
 The presence of  odd  hole can be seen as the horizontal  doughnut-like orbital carved out  from in the even-even core. The angular momentum  $\vec{j}_h$ of 
 this hole-orbital points along the $l$-axis of the core, because this orientation corresponds to the energy minimum of the short-range repulsive interaction between the hole-orbital and the core. 
The core behaves like an even-even triaxial nucleus, generating the  $\vec{R}$ component centered with the $m$- axis. 

The presence of the high-$j$ orbitals dramatically    changes the rotational behavior. As they act like gyroscopes, the nucleus becomes a clockwork of gyroscopes, which
can rotate about an axis that is tilted away from one of the principal axes of the shape. 

When there is only one particle on a high-$j$ orbital  the total angular momentum
is $\vec{J}=\vec{j}_p+\vec{R}$. At low spin  the Coriolis force exerted by particle aligns $\vec{R}$ with the $s$-axis,   and the nucleus rotates about the $s$-axis. Built on this
yrast band there are wobbling excitations with a precession cone of $\vec{R}$ about the $s$-axis. The total angular momentum $\vec{J}$ precesses around  the $s$-axis as well.
The corresponding wobbling motion of the whole nucleus in the laboratory system generates the characteristic strong $E2$ transition between the adjacent wobbling bands. 
This mode has been called transverse wobbling because the axis of the precession cone is perpendicular to the $m$-axis of maximal moment of inertia. At suffciently large spin
the larger moment of inertia of the $m$-axis overcomes the pull of the particle. The transverse mode becomes unstable and changes to longitudinal wobbling about the $m$-axis.
The analogous mechanism works for a hole added to the core, where the  $s$-axis is replaced by the $l$-axis. Both types of transverse wobbling  have been observed in a number
of odd-A nuclei and for two-quasiparticle bands in even-even nuclei. The properties of the wobbling mode have been been recently reviewed by Frauendorf \cite{Frau24}.

When in odd-odd nuclei an odd high-$j$ proton is combined with an odd  high-$j$ neutron hole (or vice versa),
the total angular momentum $\vec{J}=\vec{j}_p+\vec{j}_h+\vec{R}$, which is the axis of rotation, is tilted respect to all three principal axes of the triaxial shape.    S. Frauendorf and J. Meng \cite{FM97} first noticed that the rotating system attains chirality. Looking in Figure \ref{f:shape} from the tip of the vector $\vec{J}$ onto the shape, the $s$-, $m$- and $l$-half-axes are ordered clockwise in the right-handed configuration and counterclockwise in the left-handed configuration. Although the two configurations look like mirror images, they are {\bf not} related by a reflection as in the case of molecules. 
A reflection does not change the angular momenta ($\vec{R}=m\vec{r}\times\vec{v}$). One has to invert the direction $\vec{R}$, which selects the half-axis, in order to convert the left-handed into the right-handed configuration. (An additional rotation brings it into the position shown in the figure.) This is achieved by time reversal, which changes only the sign of the velocities. The nucleons run "backwards" ($\vec{R}=m\vec{r}\times\vec{v}$). 

The tilt of $\vec{J}$ with respect to the principal axes removes the symmetry with respect to a rotation by $\pi$ around the rotational axis, which has the consequence that all integer values of $I$
constitute a rotational band. In addition to the ones shown, there are three more left-handed and three more  right-handed configurations,  which are
generated by changing the directions of $\vec{j}_p$ and $\vec{j}_h$. They are combined into superpositions, which
 give rise to two $\Delta I=1$ bands of rotational states where the states  with the same angular momentum are expected to have
the same energy and  emit $\gamma$-radiation in the same way. The chiral partners  have the same parity quantum number, 
which indicates that the configurations do not change under reflections through the principal planes
of the triaxial shape. A more detailed discussion of the symmetries of rotating nuclei can be found in Ref. \cite{Frau01}.

There is some coupling between the left- and right-handed configurations which perturbs the ideal pattern. The energies of the chiral partners are somewhat different and so are the
$\gamma$ emission pattern. The consequences of the left-right coupling are discussed in Ref. \cite{Frau11} as part of an introduction to nuclear chirality in the style of this comment.
Chiral partner bands have been identified in the mass regions around $A= $ 80, 110,  130, 190. The evidence has been recently reviewed by Meng \cite{Meng24}.

The precession of $\vec{R}$  generates a precession of $\vec{J}$, which is shown as the ellipse in Fig. \ref{f:shape}. The corresponding lowest wobbling excitations will appear  
as two $\Delta I=1$ bands that decay via strong $E2$ transition to the chiral partner  bands. The authors of Ref. \cite{Br74} reported the first evidence for such chiral wobbling band.
The Chinese-South African collaboration studied the odd-odd nucleus $^{74}$Br  utilizing a high-quality dataset of prompt double and triple $\gamma$ coincidences following the fusion 
evaporation reaction populating the nucleus. The data were acquired with Ge Clover detectors in the AFRODITE array. Lifetimes of the excited states were determined from the
Doppler shifts of the $\gamma$ rays observed in the coincidence spectra. Absolute reduced transition  probabilities were obtained by means of angular distribution
from oriented state (ADO) ratios, linear polarization values, and the level lifetimes.

\begin{figure}[t]
\includegraphics
  [width=\linewidth]
  {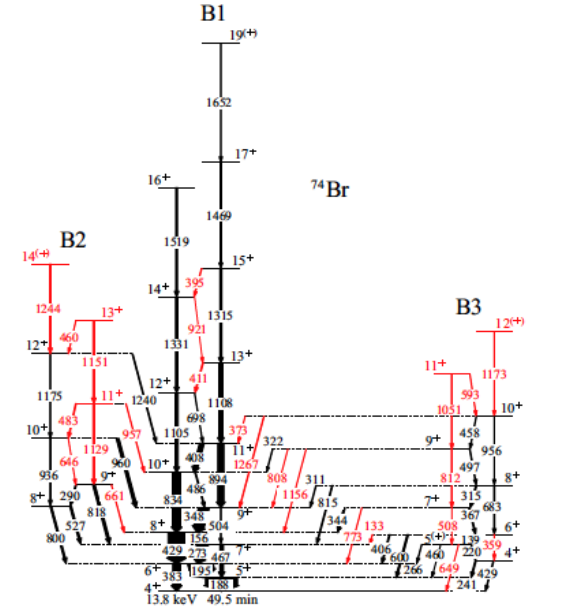}
\caption{\label{f:Br74bands} The partial level scheme for $^{74}$Br from Ref. \cite{Br74}.
The levels are arranged into the rotational bands B1 and B2, which are chiral partner bands, and B3, which represent the chiral wobbling band.
 New transitions and levels are marked as red.
The arrow width represents the intensity of the transitions. The intensities are normalized
to to the 383 keV transition as 100. Reproduced with permission from Ref. \cite{Br74}.
 }.
\end{figure}

Figure \ref{f:Br74bands} shows the new level scheme constructed by means of the cutting-edge experiment. As usual, the levels are arranged into rotational bands based 
on the strong $\Delta I=2$ $E2$ and $\Delta I=1$ mixed $E2$-$M1$ transitions. B1 constitutes the  $\Delta I=1$ yrast band characteristic for rotation about the tilted axis. 
B2 is assigned to the  $\Delta I=1$ chiral partner at somewhat higher energy. The shift reflects the left-right coupling. The assignment is supported by the dominant 
$M1$ character of the $\Delta I=1$ transitions within and between the partner bands, which stagger with opposite phase. B3 constitutes one of the chiral wobbling bands. The assignment 
 is established by the
enhanced $\Delta I=1$ $E2$ transition probabilities to B1. 

Calculation using a model which couples a triaxial rotor with a proton $h_{11/2}$ hole and a $h_{11/2}$ neutron particle  well reproduce the observed energies and transition probabilities. 
They support in a quantitive way the discussion of chiral partner bands and chiral wobbling given above.

The discovery of the chiral wobbling band  in Ref. \cite{Br74} revealed  a new facet of the exotic rotation modes of triaxial nuclei. Moreover,
the work affirmed the existence of chirality in the $A=80$ mass region.

\end{document}